\documentclass[prl,preprint]{revtex4}
\usepackage{psfig}

\begin{document}

\title{Resistance proof, folding-inhibitor drugs }
\author{R.A. Broglia$^{1,2,3}$ G. Tiana$^{1,2}$,and R. Berera$^{1}$ }
\address{$^1$Department of Physics, University of Milano,}
\address{via Celoria 16, 20133 Milano, Italy,}
\address{$^2$INFN, Sez. di Milano, Milano, Italy,}
\address{$^3$The Niels Bohr Institute, University of Copenhagen,}
\address{Bledgamsvej 17, 2100 Copenhagen, Denmark}
\date{\today}

\begin{abstract}
Conventional drugs work, as a rule, by inhibiting the enzymatic activity of specific proteins, capping their active site. In this paper we present a model of non- conventional drug design based on the inhibiting effects  small peptides obtained from segments of the protein itself have on the folding ability of the system.  Such peptides attach to the newly expressed (unfolded) protein and inhibit its folding, inhibition which cannot be avoided but through mutations which in any case denaturate the enzyme. These peptides, or their mimetic molecules, can be used as effective alternative drugs to those already available, displaying the advantage of not suffering from the upraise of resistence.
\end{abstract}

\maketitle

Drugs perform their activity either by activating or by inhibiting some 
target component of the cell. In particular, many inhibitory drugs 
bind to an enzyme and deplete its function 
by preventing the binding of the substrate. This is done by either 
capping the active 
site of the enzyme (competitive inhibition) or, binding to some 
other part of the
enzyme, 
by provoking structural changes which make the enzyme unfit to 
bind the substrate 
(allosteric inhibition).
The two main features that inhibitory drugs must display  are efficiency 
and specificity. In fact, it is not sufficient that the drug binds to 
the substrate and reduces efficiently 
its activity. It is also important that it does not interfere with 
other cellular processes, 
binding only to the protein it was designed for. These features are 
usually accomplished 
designing drugs which mimick the molecular properties of the natural 
substrate. 
In fact, 
the pair enzyme/substrate have undergone milions of years of evolution in 
order to display 
the required features. Consequently, the more similar the drug is to 
the substrate, 
the lower is the probability that it interferes with other cellular processes.
Something that this kind of inhibitory drugs are not able to  do is to avoid the development 
of resistance, a phenomenon which is typically related to viral protein targets. 
Under the selective pressure of the drug, the target is often able to either 
mutate the 
amino acids at the active site or at sites controlling its conformation in such a way 
that the activity of the enzyme is 
essentially retained, while the drug is no longer able to bind to it. An important 
example of drug-resistance is connected with AIDS. In this case, one of the
main  target proteins, HIV-protease, is able to mutate its active site so as
to avoid the effects of drug action within a period of time of 6-8 months
(cf. e.g. \cite{new1,new2}).
In the present paper we discuss the design of drugs which interfere 
with the folding mechanism of the target protein, destabilizing it and making 
it prone 
to proteolisis. We shall show that these drugs  are efficient, specific and
do not suffer from the 
upraise of resistance.
The model of protein folding we employ is largely used in the
literature. In spite of its 
simplicity, it reproduces well the thermodynamic and kinetic properties of real
proteins \cite{sh1,sh2,jchemphys1}. The model describes a protein as a chain of beads sitting 
on a cubic lattice, each bead representing an amino acid and interacting with the 
neighbouring beads through a contact potential. There are twenty kinds of beads to 
account for the twenty kinds of natural amino acids. Consequently the contact 
potential is defined by a 20$\times$20 matrix, extracted from statistical 
analysis of the contacts of real proteins \cite{mj}.

Using this model, it has been shown that single domain proteins fold according to a hierarchical 
mechanism \cite{jchemphys2,jchemphys3}. Starting from an elongated conformation,
it is found that, highly conserved and strongly interacting amino-acids
lying close along the designed chain form small local elementary structures
(LES).
Due to the small conformational space available and to their 
large attractive propensity, these LES are formed at a very early stage in the
folding process and are very stable. The rate limiting step
of this process corresponds to the assembly of the LES 
to build their native, non--local contacts(folding nucleus). This nucleation can be 
done in a relatively short time, because LES, moving as almost
rigid entities, not only reduce the conformational space available to the
protein but also display low probability of forming non-native
interactions. Furthermore they interact with each other 
more strongly than single amino acids belonging to these structures do\cite{jchemphys3}. 
The nucleation event corresponds to the overcoming of the major free energy barrier
found in the whole folding process \cite{sh_nucl}. 
After this is accomplished the remaining conformational space available to
the protein is so small that the
system reaches the native state almost immediately. 
In keeping with these results we suggest the use of short peptides with the same sequence as
the
LES (in the following, shortened as p--LES) to destabilize the protein. 
We test this suggestion on three sequences designed to fold to the three different structures 
displayed in Fig. 1. The corresponding sequences are listed in the caption
to the figure. 
It was shown in a previous work \cite{1d3d} that the associated LES are 
built out of residues 3--6, 11--14 and 27--30 for sequence (a) of Fig.
1 (known as S36 in the literature), of 
residues 1--6, 20--22 and 30--31 for sequence (b) and of residues 34--42 and 2--12 for sequence (c).

To asses the ability p--LES display in destabilizing designed proteins, we have performed Monte Carlo simulations 
of a system composed of the protein and a number $n_p$ of p--LES in a cubic cell of linear size $L$ 
with periodic boundary conditions \cite{dimer}. Each simulation starts from a random conformation of 
the system and is carried on through $10^8$ MC steps at fixed temperature $T$. During the simulation, 
we have collected the histogram of the order parameter $q$, defined as the
relative number 
of native contacts, parameter which measures the extent to which the equilibrium
state reached by the protein is similar to the native conformation.  

In Fig. 2(a) we display the equilibrium distribution of $q$, calulated at $T=0.24$ and  $L=7$ 
for the system composed of sequence S36 and a number of p--LES 3'--6' as a function of $n_p$ (concentration)\cite{footnote}. While the distribution of $q$ values in
the absence of p--LES (solid line) 
shows a two--peaks shape, reflecting a all-none transition between the native ($q>0.7$) 
and the unfolded ($q<0.6$) state, the presence of p--LES reduces markedly the stability of 
the protein. The effect of the other p--LES, i.e. 11'--14' and 27'--30', is similar
to that found for the peptide 3'--6'and
is 
displayed in Figs. 2(b) and (c). The strength of the inhibitory effect, measured in terms of relative 
population $p_{1}$ of the native state ($q>0.7$) of the protein in presence of  p--LES, is displayed in Fig. 3.

To further test the validity of these results, we have repeated the above calculations making use of peptides 
corresponding to segments of the protein sequence other than those
corresponding to LES. 
In Fig. 2(d) the effect of peptides corresponding to residues 8--11 is shown. One can notice 
that the protein is not destabilized to any significant extent. To ensure that this
result is not a consequence of the  weak binding of the peptide to the protein, 
we have replaced the amino-acids  8'--11' of the peptide by amino acids
which interact with the complementary amino acids of the protein (i.e. amino
acids 21,22,15 and 14 respectively, cf.Fig1(a)) as strongly as those
belonging to LES do.
No difference with the results shown in Fig.2(d) was found.

The thermodynamics which is at the basis of the disruptive mechanism of p--LES is
quite simple. In fact 
there are three thermodynamically relevant states in the range of temperatures where the 
protein is stable: 1) the state in which the protein is folded and the $n_p$ p--LES do not 
interact with the protein (whose free energy is taken as reference and assigned a value $\Delta
F_1=0$), 
2) the state in which a p--LES is bound to the (complementary) LES of the protein 
preventing it from folding, its free energy being $\Delta F_2=\Delta F_0+E_{LES}+TS_t-T\log
n_p$ where the 
 quantity $\Delta F_0$ is the difference in free energy between the unfolded and the native state of 
the isolated protein, $E_{LES}$ is the interaction energy between the p--LES and the
complementary 
LES and $S_t$ is the translational entropy of a p--LES, 3) the state in which the protein is 
unfolded and the p--LES do not interact with the protein, the associated free energy being $\Delta F_3=\Delta F_0$. 
The translational entropy can be estimated using the relation
\begin{equation}
S_t(n_p)=\log\left[12\cdot(V-v_{prot}-(n_p-1)v_{ples})\right],
\end{equation}
where $V=343$ is the volume of the cell (in lattice units) in which the simulations are performed, 
$v_{prot}=166$ is the average volume occupied by the protein, $v_{ples}=20$ is the average volume 
occupyied by a p--LES, while the prefactor $12$ accounts for the orientation of the p--LES. It then follows that the 
equilibrium probability that the protein is folded, i.e. in state 1), is given by
\begin{equation}
\label{p1}
p_1=\frac{1}{1+e^{-\Delta F_0/T}[n_p e^{-S_t(n_p)-E_{LES}/T}+1]}.
\end{equation}

In Figs. 3(a), (b) and (c) are displayed the values of $p_1$ associated with
the sequence $S36$ and the three 
p--LES 3'--6', 11'--14' and 27'--30' as a function of $n_p$ (solid dots). The
continuous curves 
are the the results obtained making use of  Eq. (\ref{p1}) and of the
numeric values 
$\Delta F_0=-0.038$ (obtained from MC simulations) and $E_{LES}$
($=-2.5$(a),$-2.0$(b),-2.5(c)).

The overall agreement found between the three-state model and the results of
MC simulations suggests that the destabilization 
of the protein is, in fact, due to the binding of the p--LES to the protein. 
Naively speaking, the protein prefers to bind the p--LES instead of the native LES because in 
this ways it saves internal entropy, which is not compensated by the loss of translational
entropy. 

The above results also suggest that the state associated with the p--LES bound to sites of the 
protein surface different from the LES is not relevant. This state has a relative free energy 
 $\Delta F_4=E'+TS_t-T\log n_p$, where $E'$ is the interaction energy between the p--LES and 
the surface of the protein. The effect of this state on the stability of the protein would be 
to raise the asymptotic value of $p_1$ for large values of  $n_p$. The fact that in none of the cases  
studied (cf. Figs. 3(a), (b) and (c) ) the asymptotic value of $p_1$ is different from zero
indicates 
that $E'\ll E_{LES}$. In other words, the binding properties of the p--LES are highly specific. 
It could hardly have been different: since p--LES are identical to LES, a propensity of LES to bind 
some non--native part of the protein would imply the stabilization of a metastable state, 
something that evoultion tends to avoid.

In Fig. 3(d) we display the dependence of $p_1$ with temperature (solid dots) 
for the case of p--LES 3'--6'and $n_{p}=2$.  The results of the
simulations are well reproduced by the predictions obtained making use of Eq.
(\ref{p1}). 
In these estimates the temperature dependence of $\Delta F_0$  has been approximated with that 
of the Random Energy Model \cite{derrida,jchemphys2}. The non--monotonic behaviour of $p_1(T)$ is 
a consequence of the competition between the stabilization of the native state and the decrease of 
the free energy of the unfolded states taking place as the temperature is lowered. At high 
temperatures, the state 3), which is independent on $n_p$, becomes
important, weakening the overall dependence 
of $p_1$ on $n_p$.
We have repeated the calculations described above, but this time making use
of sequences obtained from S36 by introducing random point mutations in the
LES 27-30. In this way we try to mimic the development of drug resistance of
a viral protein. We observe two situations: I)if the protein is (upon
mutation) still able to fold the scenario corresponding to Figs 1(a)-(c) is
still valid, II)if the mutation denaturates the protein, the p-LES does not,
essentially, bind any more to it.

We have found that also the dynamical properties of p--LES make them suitable to be used as drugs.  Starting 
from a random conformation of the protein and of the peptides, we have 
calculated the probability $P(t)$ that the bond between residue 30 of the protein and 3' of any of the p--LES 3'-6' is formed as a function of time. 
This bond is chosen as representative of the interaction between the whole LES
27--30 
and the p--LES 3'-6', the dynamics of the other bonds associated with  the same LES being quite
similar. The shape of the calculated probability function is well 
fitted by a single exponential $P(t)\sim (1-\exp(-t/\tau'))$, where $\tau'$ is 
the characteristic time of bond formation. The dependence of $\tau'$ on 
the number $n_p$ of p--LES is displayed in Fig. 4 as a solid line, where it is compared to 
the average time needed for the p--LES to build the bond 30-3' with the protein after 
a random search in the volume of the cell, that is
\begin{equation}
\frac{12\cdot(V-v_{prot}-(n_p-1)v_{ples})}{n_p}.
\end{equation}
The result obtained making use of this relation is also displayed in Fig. 4
(dashed curve). The agreement with the result of the numerical simulations 
indicates that the random search is the actual mechanism which leads to the binding
of the p-LES to its (complementary LES. The 
fact that $P(t)$ is well reproduced by a single exponential indicates
furthermore that this is the only mechanism
operative. In particular, this result  excludes the possibility that the p--LES binds tightly 
to some other 
part of the protein. Such a scenario  would produce a double-- or more--fold--exponential 
shape of $P(t)$.  

To be noted that the binding time $\tau'$ of p--LES to the protein is much shorter than the binding 
time of the associated native contact between LES within the protein. 
In particular, the result displayed  Fig. 4 and associated with contact
30-3'
is to be compared to the 
value $\tau=1.3\cdot 10^5$ of the native contact 30-3 \cite{jchemphys3}. The reason for
this result is associated with the fact that, 
unlike LES, p--LES are not slowed down by the polymeric connection with 
the rest of the protein. A consequence of this fact is the ability  p--LES 
have to bind to LES of the protein even if this is in its (equilibrium) native state. 
The p--LES can take advantage of the thermal fluctuations of the protein and 
make use of the fact that these fluctuations display a recursion time 
(which, assuming that the system is ergodic, is equal to the mean first 
passage time) much longer than the time needed by p--LES  
to enter and disrupt the protein by binding to one of its the LES. As a matter of fact, we
have calculated the 
distribution of $q-values$ starting from the protein in the native state, 
finding the same distribution as that displayed in Fig. 2.

Calculations as those described above and leading to the results displayed
in Figs. 2-4 have also been carried out for the other two model proteins 
displayed in Fig. 1.
The outcome of these calculations are, as a rule, in agreement with those found in
connection with sequence S36. To be noted, however, an important difference
found in 
connection with sequence b) (36-mer). This designed protein displays, in the folding
process, three LES of length 2, 3 and 6, respectively. 
While the p--LES built of 6 residues inhibits folding as those described above, 
the other two p--LES do not. This is connected with the small size of 
these p--LES, which makes them quite unspecific. In fact, the probability that a p--LES 
binds to some part of the protein other than the target LES decreases exponentially 
with the number of residues involved.

We have shown that it is possible to inhibit the activity of a protein by
disrupting its folding with the help of small peptides which mimick the 
LES of the protein. 
The very reason why LES make single domain proteins fold fast confers p--LES the required features
to act as  effective drugs, 
that is, efficiency and specificity. They are efficient because they bind as strongly as
LES do. 
Since LES are responsible for the stability of the protein, their stabilization energy 
must be of the order of several times $kT$. These peptides are also as specific as LES
are. 
In fact LES have evolved so as to prevent the upraise of metastable states and
to avoid aggregation, aside of securing the 
protein to fold fast. 
The possibility of developing non--conventional drugs for
actual situations is tantamount to being able to determine the LES for a given protein. 
This can be done either experimentally 
(e.g. making use of $\varphi$--value analysis\cite{fersht} or ultrafast stopped 
flow experiments) or extending the algorithm discussed in ref. \cite{1d3d}
making use of a 
realistic force field. The resulting peptides can be used either directly as drugs, or 
as templates to build mimetic molecules, which eventually do not display
side effects connected 
with digestion or allergies.

A feature which makes, in principle, these drugs quite promising as compared
to conventional ones is to be found in the fact that the
target protein cannot evolve through mutations to escape the drug, as
happens in particular in  the case of viral 
proteins, because the mutation of residues in the LES would anyway lead to
protein denaturation.



\begin{figure}
\centerline{\psfig{file=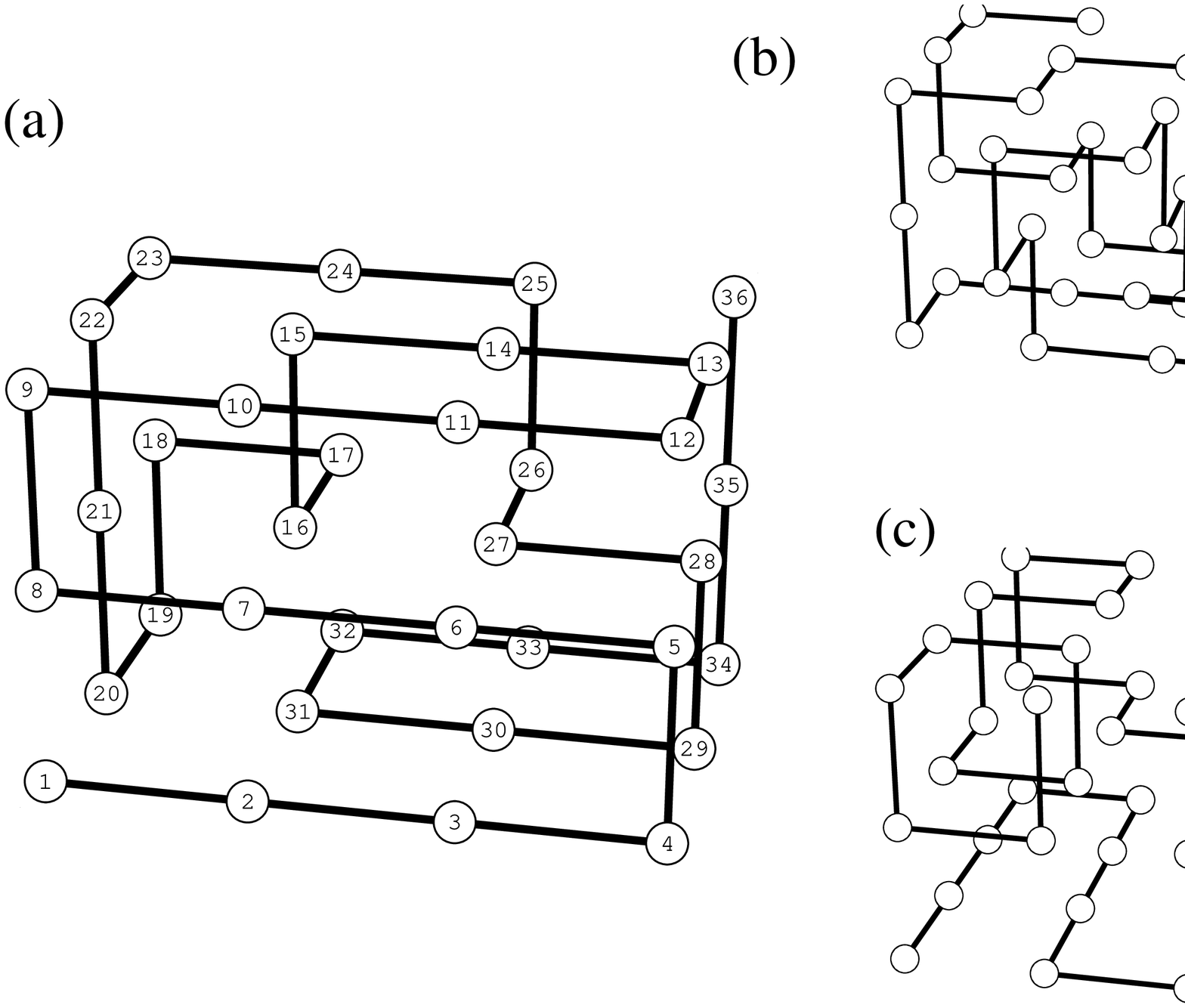,height=5cm,width=8cm}}
\caption{The three native structures used in the present calculations (two
36-mers and one 48-mer). The associated designed sequences 
\protect\cite{jchemphys2} are (a) S36$\equiv$SQK\-WLE\-RGA\-TRI\-ADG\-DLP\-VNG\-TYF\-SCK\-IM\-ENV\-HPLA, (b) RASM\-KDKTV\-GIGHQ\-LYLNFE\-GEWCPA\-PDN\-TRV\-SLAI, (c) IMES\-QKW\-LCM\-EPAH\-WCVY\-TIQG\-LGNV\-NCPN\-TREF\-DSGR\-SKIQ\-DAY\-LFH. }
\end{figure}

\begin{figure}
\centerline{\psfig{file=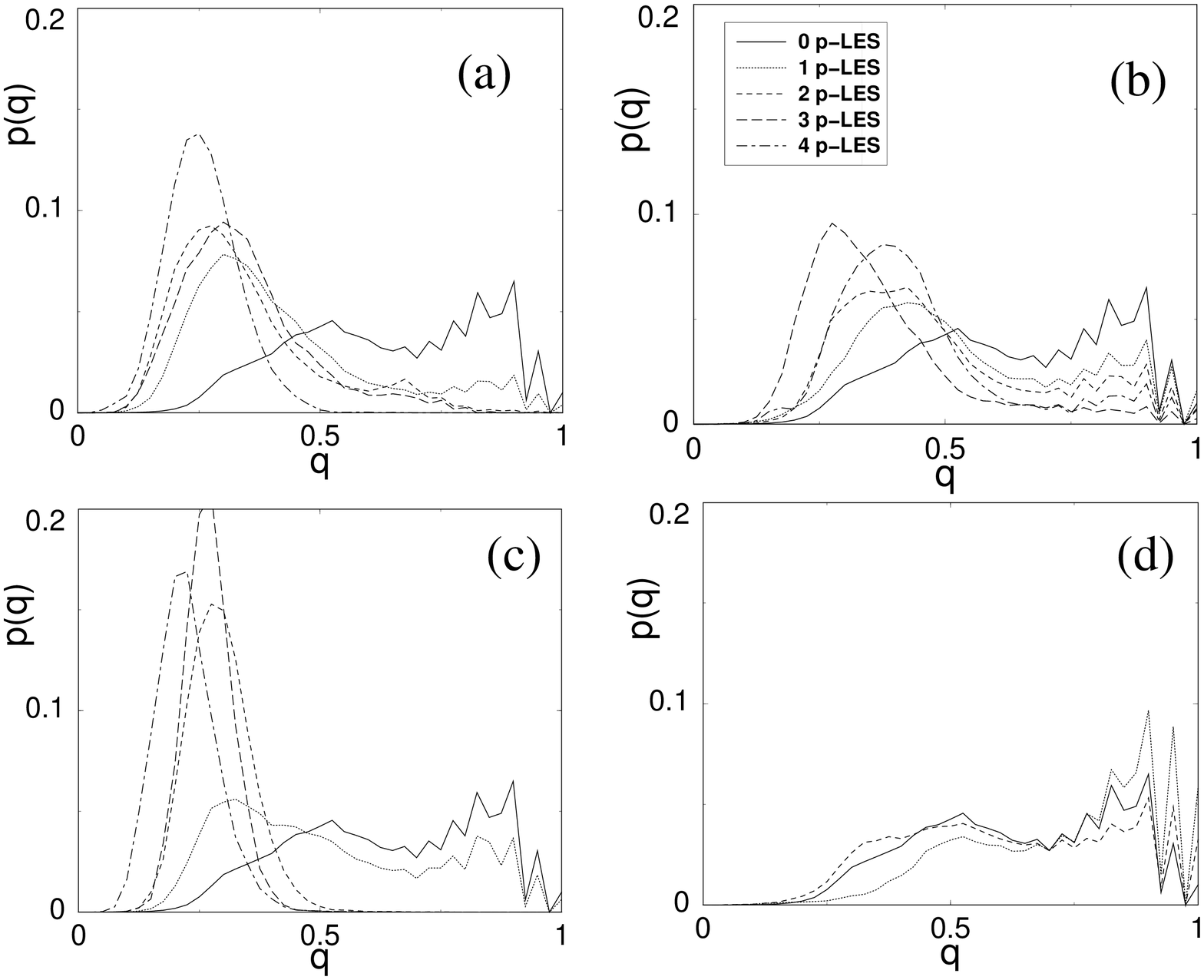,height=5cm,width=8cm}}
\caption{The equilibrium distribution of the order parameter $q$ of sequence (a)
(cf. caption to Fig. 1) 
in presence of $n_p$ p--LES of kind 3'--6' (a), 11'--14' (b) and 27'--30'
(c), calculated at temperature T=0.24 in the units chosen
(R$T_{room}=0.6\;kcal/mol$).  As
control, a string corresponding to the residues 8--11 of the protein was
also used (d).}
\end{figure}

\begin{figure}
\centerline{\psfig{file=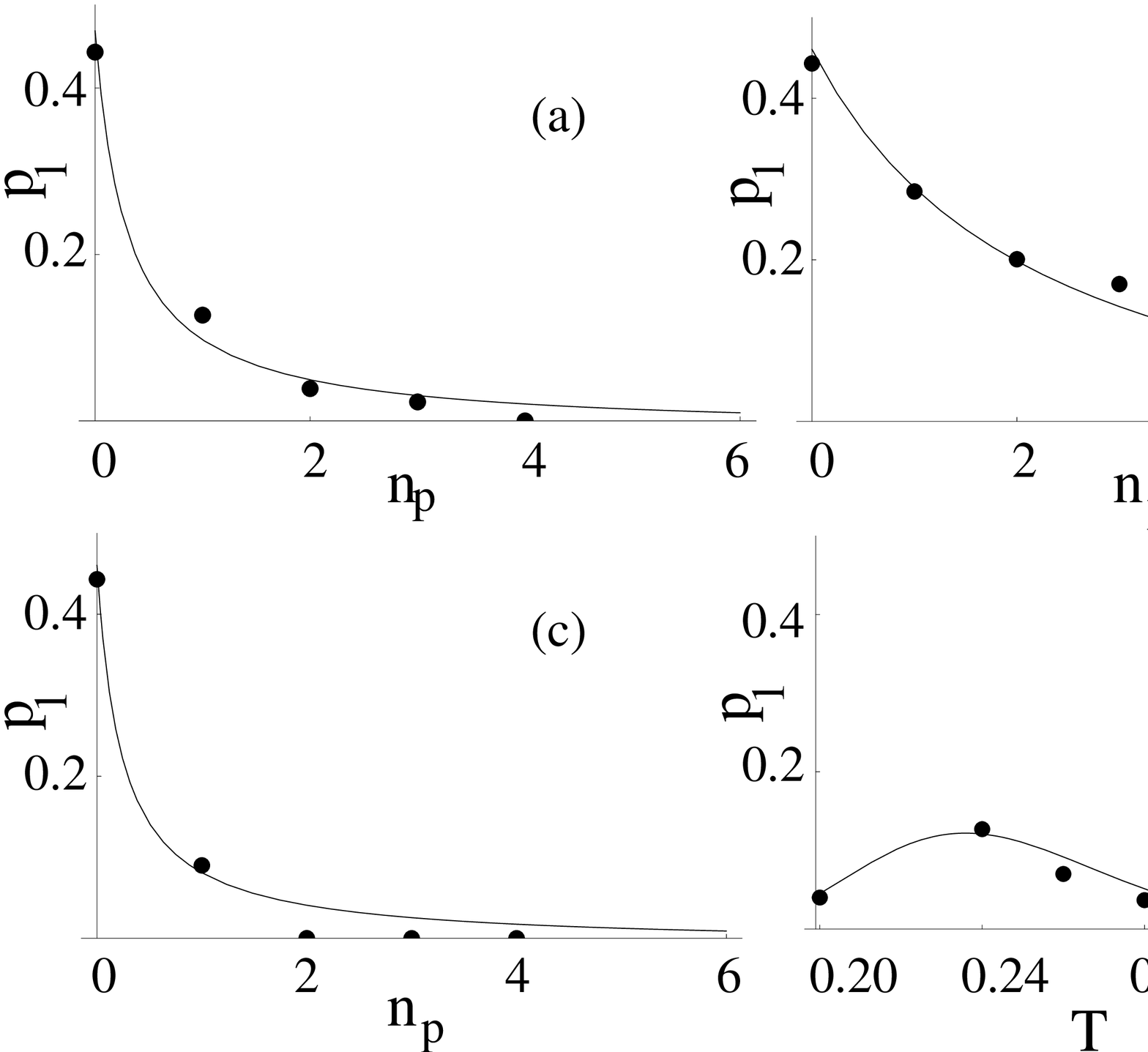,height=5cm,width=8cm}}
\caption{Stability $p_1$ (for T=0.24) of the native structure of S36
(protein (a) of Fig. 1) as a function 
of the number $n_p$ of p--LES of kind 3'--6' (a), 11'--14' (b) and 27'--30' (c) present 
in the cubic cell (solid dots). The results displayed by the continuous
curve  was determined
making use  of Eq. (2) as discussed in the text.(d) The quantity $p_1$ associated with $n_{p}$=2
p-LES 3'-6'
as a function of temperature.}
\end{figure}

\begin{figure}
\centerline{\psfig{file=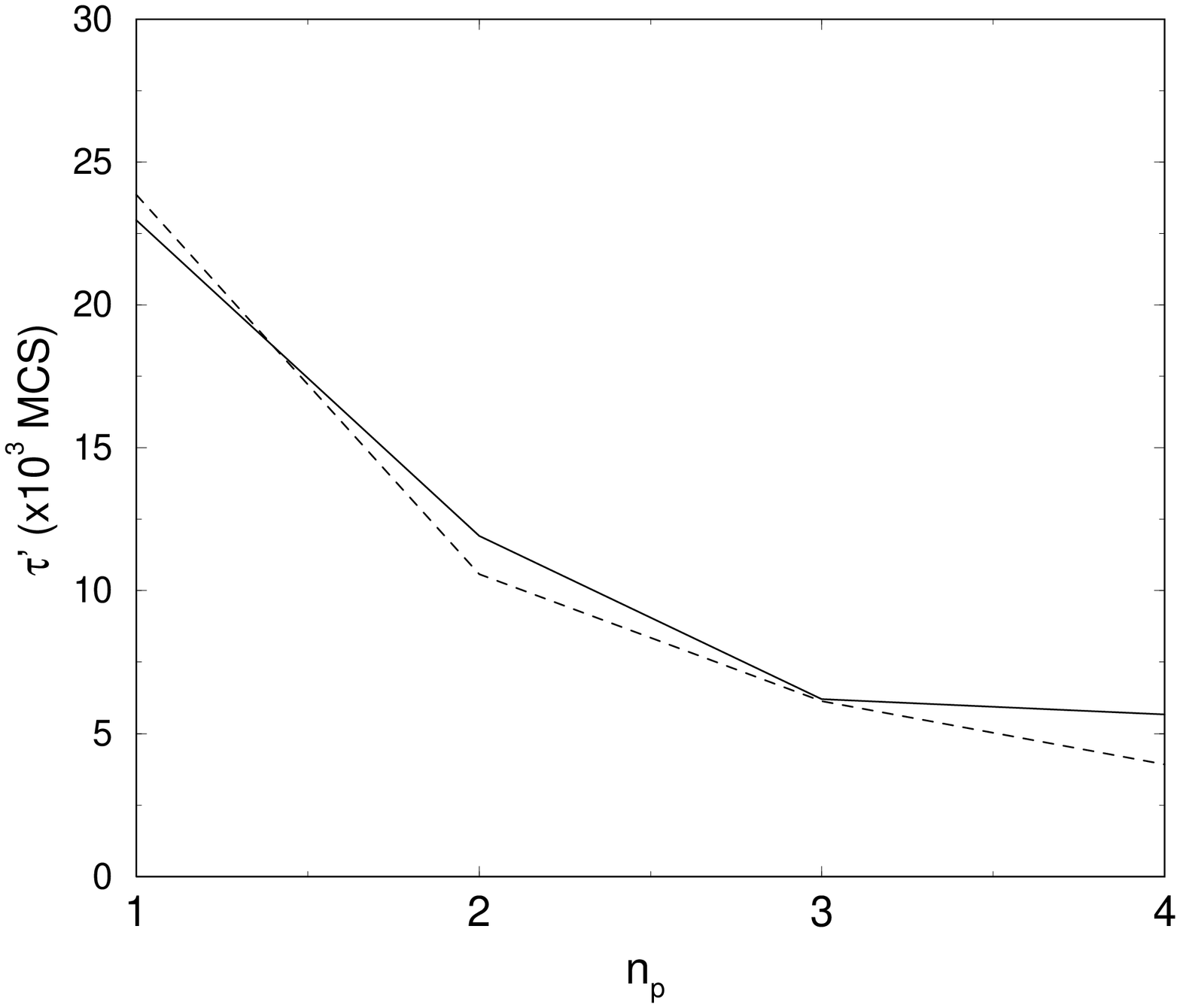,height=5cm,width=8cm,angle=-90}}
\caption{The mean binding time $\tau'$ between the residue 30 of the protein and 3' of the p--LES, 
as a function of the number $n_p$ of
p--LES (solid line). The result of MC simulations is compared with the random search 
time predicted making use of Eq. (3) (dashed line).}
\end{figure}


\begin{thebibliography}{99}
\bibitem{footnote} We 
indicate with primed numbers the residues belonging to a p--LES, following the numeration 
of the protein sequence. Thus, 3'--6' means the p--LES built out of four
monomers corresponding 
to residues 3--6 of the protein.
\end{thebibliography}
\end{document}